# Enabling propagation of anisotropic polaritons along forbidden directions via a topological transition

Jiahua Duan[1,2†], Gonzalo Álvarez-Pérez[1,2†], Kirill V. Voronin[3], Iván Prieto[4], Javier Taboada-Gutiérrez[1,2], Valentyn S. Volkov[3], Javier Martín-Sánchez[1,2*], Alexey Y. Nikitin[5,6*], Pablo Alonso-González[1,2*]

[1]Department of Physics, University of Oviedo, Oviedo 33006, Spain.
[2]Center of Research on Nanomaterials and Nanotechnology, CINN (CSIC-Universidad de Oviedo), El Entrego 33940, Spain.
[3]Center for Photonics and 2D Materials, Moscow Institute of Physics and Technology, Dolgoprudny 141700, Russia.
[4]Institute of Science and Technology Austria, am Campus 1, Klosterneuburg 3400, Austria.
[5]Donostia International Physics Center (DIPC), Donostia/San Sebastián 20018, Spain
[6]IKERBASQUE, Basque Foundation for Science, Bilbao 48013, Spain.
* pabloalonso@uniovi.es, alexey@dipc.org, javiermartin@uniovi.es
† These authors contributed equally to this work.

**Recent discoveries of polaritons in van der Waals (vdW) crystals with directional in-plane propagation, ultra-low losses, and broad spectral tunability have opened the door for unprecedented manipulation of the flow of light at the nanoscale. However, despite their extraordinary potential for nano-optics, these unique polaritons also present an important limitation: their directional propagation is intrinsically determined by the crystal structure of the host material, which imposes forbidden directions of propagation and hinders its control. Here, we theoretically predict and experimentally demonstrate that directional polaritons (in-plane hyperbolic phonon polaritons) in a vdW biaxial slab (α-MoO$_3$) can be steered along previously forbidden directions by inducing an optical topological transition, which naturally emerges when placing the slab on a substrate with a given negative permittivity (4H-SiC). Importantly, due to the low-loss nature of this topological transition, we are able to visualize in real space exotic intermediate polaritonic states between mutually orthogonal hyperbolic regimes, which permit to unveil the unique topological origin of the transition. This work provides new insights into the emergence of low-loss optical topological transitions in vdW crystals, offering a novel route to efficiently steer the flow of energy at the nanoscale.**

Phonon polaritons (PhPs) —hybrid electromagnetic waves arising from the coupling of photons and optical phonons in polar dielectrics— in materials consisting of atomic planes bonded by weak vdW forces (vdW materials) hold great promises for infrared nanophotonics[1-3], as they feature deep-subwavelength confinement and ultralow-loss propagation. Particularly, PhPs in so-called hyperbolic media (HPhPs) exhibit extremely anisotropic dispersion: that is, their iso-frequency curve (IFC), a slice of the dispersion in momentum-frequency space by a plane of a constant frequency $\omega_0$, describes a hyperbola-like curve[4-9]. Consequently, the propagation of HPhPs is strongly anisotropic, manifesting itself in a very peculiar fashion. For instance, HPhPs launched by a point source feature concave wavefronts and strong directionality due to the enormous density of optical states at specific wavevectors (along the asymptotes of the IFC). This exotic polaritonic propagation was first reported for out-of-plane HPhPs in the uniaxial crystal hexagonal boron nitride[10-12] (h-BN), showing remarkable potential for applications in photonics[6], specifically for sub-diffractional focusing[13,14] and sensing via strong coupling with organic molecules[15]. More recently, in-plane HPhPs have been discovered in the biaxial vdW crystals α-phase molybdenum trioxide[16,17] (α-MoO$_3$) and α-phase vanadium pentaoxide[18] (α-V$_2$O$_5$), opening the door to study directly on the crystal surface extraordinary optical phenomena stemming from the low-loss directional propagation of HPhPs. However, the direction of propagation of HPhPs in naturally hyperbolic media is difficult to control, since it is predefined by the intrinsic anisotropy of the crystal, which also imposes forbidden directions of propagation. This intrinsic limitation reduces the potential and versatility of HPhPs for applications requiring specific propagation directions of the electromagnetic energy flow at the nanoscale, such as coupling between quantum emitters[19,20] or heat management[21]. Specifically, in-plane hyperbolic polaritons can only propagate along the directions between the asymptotes of their hyperbolic IFC in the $(k_x, k_y)$ plane, i.e. within the sectors $|\tan(k_x/k_y)| < \sqrt{(-\varepsilon_y/\varepsilon_x)}$, where $\varepsilon_x$ and $\varepsilon_y$ are the in-plane permittivities of the hyperbolic medium and $k_x$ and $k_y$ are the in-plane components of the wavevector. Therefore, any engineering, manipulation or reorientation of the IFC will obviously have a dramatic effect on the directional propagation of hyperbolic polaritons in real space. Recently, attempts to enhance the anisotropic propagation of polaritons in vdW crystals have been carried out by either assembling crystal heterostructures[22], or fabricating metasurfaces[23] and twisted bilayers[24-27], which have even shown the possibility of inducing the propagation of HPhPs along one specific direction. However, despite this extraordinary progress, a solution to overcome the existence of forbidden directions of propagation of HPhPs has remained elusive so far, even from a theoretical point of view. Here, we theoretically predict and experimentally demonstrate that reorientation of the IFC in a biaxial slab can be engineered to allow directional propagation of hyperbolic polaritons along forbidden directions by inducing a low-loss optical topological transition, which naturally emerges when the biaxial slab is placed on a substrate with a given negative permittivity. In particular, we demonstrate that the direction of propagation of in-plane HPhPs in a slab of α-MoO₃ can be rotated by 90° (and thus occur along previously forbidden directions) by placing the slab on a 4H-SiC substrate. Interestingly, this rotation results from a low-loss topological transition that allows to experimentally reveal unique polaritonic intermediate states yielding propagation of HPhPs along mutually orthogonal directions. By theoretically analyzing these states, we unveil the unique topological origin of this transition, characterized by a gap opening in the polaritonic dispersion.

To elaborate on the theoretical conditions upon which such reorientation of the directional propagation of HPhPs can take place, let us start with the general case of a biaxial slab embedded between two semi-infinite isotropic media with dielectric permittivities $\varepsilon_1$ and $\varepsilon_3$. Without loss of generality, we assume the representative dielectric functions $\text{Re}(\varepsilon_x) < 0$, $\text{Re}(\varepsilon_y)$, $\text{Re}(\varepsilon_z) > 0$, in analogy with biaxial vdW materials in which in-plane HPhPs have been reported to propagate within hyperbolic sectors centered along the direction with negative permittivity[16-18,28], i.e. the $x$-axis, such as α-MoO₃ and α-V2O5. The dispersion relation of polaritons in a biaxial slab (considering the layers stacked along the $z$ axis) reads[29]:

$$k(\omega) = \frac{\rho}{d}\left[\arctan\left(\frac{\varepsilon_1 \rho}{\varepsilon_z}\right) + \arctan\left(\frac{\varepsilon_3 \rho}{\varepsilon_z}\right) + \pi l\right], \quad l = 0,1,2 \dots, \quad (1)$$

where $d$ is the thickness of the biaxial slab, $\boldsymbol{k}$ is the in-plane momentum ($k^2 = k_x^2 + k_y^2$), $\alpha$ is the angle between the $x$ axis and $\boldsymbol{k}$, $\rho = i\sqrt{(\varepsilon_z/(\varepsilon_x \cos^2\alpha + \varepsilon_y \sin^2\alpha))}$, and $l$ is the mode index. Setting in Eq. (1) $k_y = 0$ (or $\alpha = 0$), we find that the previously reported propagation of in-plane hyperbolic polaritons centered along the $x$ axis is allowed when $\text{Re}(\varepsilon_1) + \text{Re}(\varepsilon_3) > 0$ (see details in Supplementary Information). Importantly, we also find that propagation of hyperbolic polaritons is allowed along the dielectric $y$ axis, initially a non-polaritonic direction[16,17,27], —i.e. physical solutions of Eq. (1) for $\alpha = \pi/2$— when the condition $\text{Re}(\varepsilon_1) + \text{Re}(\varepsilon_3) < 0$ is fulfilled (Supplementary Information). Therefore, these results predict a 90°-reorientation of the hyperbolic IFC of polaritons in a biaxial slab in the neighborhood of $\text{Re}(\varepsilon_1) + \text{Re}(\varepsilon_3) = 0$.

Remarkably, this prediction can be addressed by using natural materials, such as the biaxial vdW semiconductor α-MoO3 placed on top of the polar crystal 4H-SiC, since α-MoO3 supports low-loss in-plane HPhPs in its *reststrahlen* band[27] (821 – 963 cm$^{-1}$) along which 4H-SiC has negative permittivity[30] (Re($\varepsilon_3$) < 0 from 797 to 970 cm$^{-1}$). Importantly, the permittivity of 4H-SiC attains a value Re($\varepsilon_3$) = −1 at the surface optical (SO) phonon frequency $\omega_{SO}$ = 943 cm-1 (Supplementary Information), where it thus approximately fulfils the critical condition derived for the IFC rotation, Re($\varepsilon_1$) + Re($\varepsilon_3$) = 0, when air is considered as superstrate ($\varepsilon_1$ = 1). To explore this possibility, we compare the analytical dispersion of HPhPs in a slab of α-MoO3 when it is placed on top of a 4H-SiC substrate and on top of a commonly used dielectric substrate, such as SiO2, where the condition Re($\varepsilon_1$) + Re($\varepsilon_3$) > 0 is always fulfilled. Figure 1a shows the dispersion of propagating HPhPs, calculated analytically from Eq. (1) (solid line) and by full-wave numerical simulations (dots), along both in-plane crystal directions ([100] and [001], corresponding to the $x$ and $y$ axes previously considered) of a 60-nm-thick α-MoO3 slab placed on top of a SiO2 substrate. In the *reststrahlen* band of α-MoO3 and below the longitudinal optical (LO) phonon frequency[28], $\omega_{LO}$ = 963 cm$^{-1}$ (Fig. 1a), the α-MoO3/SiO2 heterostructure supports the lowest-order polaritonic mode ($l$ = 0) along the [100] crystal direction of α-MoO3, i.e. along the crystal direction with negative permittivity, while there are no polaritonic modes along the dielectric [001] direction, thus constituting a forbidden propagation direction. As such, the $l$ = 0 polaritonic mode exhibits in-plane hyperbolic dispersion, consistently with what has been reported in prior works[16,17]. Such hyperbolic dispersion is more clearly visualized by plotting the IFC at a representative incident frequency $\omega_O$ = 934 cm$^{-1}$ (Fig. 1b). In this case, the IFC consists of a closed bow-tie-shaped curve composed of a finite hyperbola whose major axis lays along $k_{[100]}$, describing propagating modes for which Re($\mathbf{k}$) > Im($\mathbf{k}$) (blue lines in Fig. 1b). Note that the IFC passes through the origin, where the modes are overdamped, fulfilling Im($\mathbf{k}$) > Re($\mathbf{k}$) (grey lines in Fig. 1b), the latter regime often overlooked in the literature. As such, the IFC gives rise to propagation of HPhPs within hyperbolic sectors centered along the [100] crystal direction (Fig. 1c). Interestingly, these modes present a slab- or volume-confined propagation, as shown by extracting cross-sections of both the near-field distribution Re($E_z$) (inset of Fig. 1c) and the norm of the Poynting vector $|\mathbf{S}|$ (Supplementary Information) across the blue-shaded section in Fig. 1c.

In contrast, when the α-MoO3 slab is placed on top of 4H-SiC (with air as superstrate, $\varepsilon_1$ = 1), the polaritonic dispersion changes dramatically (Fig. 1d). The lowest-order hyperbolic mode $l$ = 0 shows now dispersion along either the [100] or the [001] α-MoO3 crystal directions (blue line and dots on the left and right panels of Fig. 1d, respectively), depending on the excitation frequency. Particularly, along the [100] direction, the $l$ = 0 mode is observed between the surface optical phonon of 4H-SiC, $\omega_{SO}$ = 943 cm$^{-1}$ and the LO phonon of α-MoO3, $\omega_{LO}$ = 963 cm$^{-1}$, i.e. when Re($\varepsilon_3$) > −1. On the other hand, it shows dispersion along the [001] direction below $\omega_{SO}$, i.e. when Re($\varepsilon_3$) < −1, where it shows negative phase velocity, $v_p = \omega/k < 0$. The corresponding IFC of the $l$ = 0 mode at $\omega_O$ = 934 cm$^{-1}$ in the α-MoO3/4H-SiC heterostructure consists of a closed bow-tie-shaped curve composed of a finite hyperbola-like curve for propagating modes with its major axis laying now along $k_{[001]}$. The emergence of available wavevectors in the mode dispersion along the $k_{[001]}$ axis leads to propagation of HPhPs within hyperbolic sectors centered along the previously non-polaritonic [001] α-MoO3 crystal direction (Fig. 1f). Such reorientation of the IFC clearly illustrates a 90º rotation of the propagation direction of HPhPs in the α-MoO3/4H-SiC heterostructure. Interestingly, HPhPs along the [001] α-MoO3 crystal direction show surface-confined propagation, as observed by extracting cross-sections of the near-field distribution Re($E_z$) (inset of Fig. 1f) and the norm of the Poynting vector $|\mathbf{S}|$ (Supplementary Information) across the blue-shaded section in Fig. 1f.

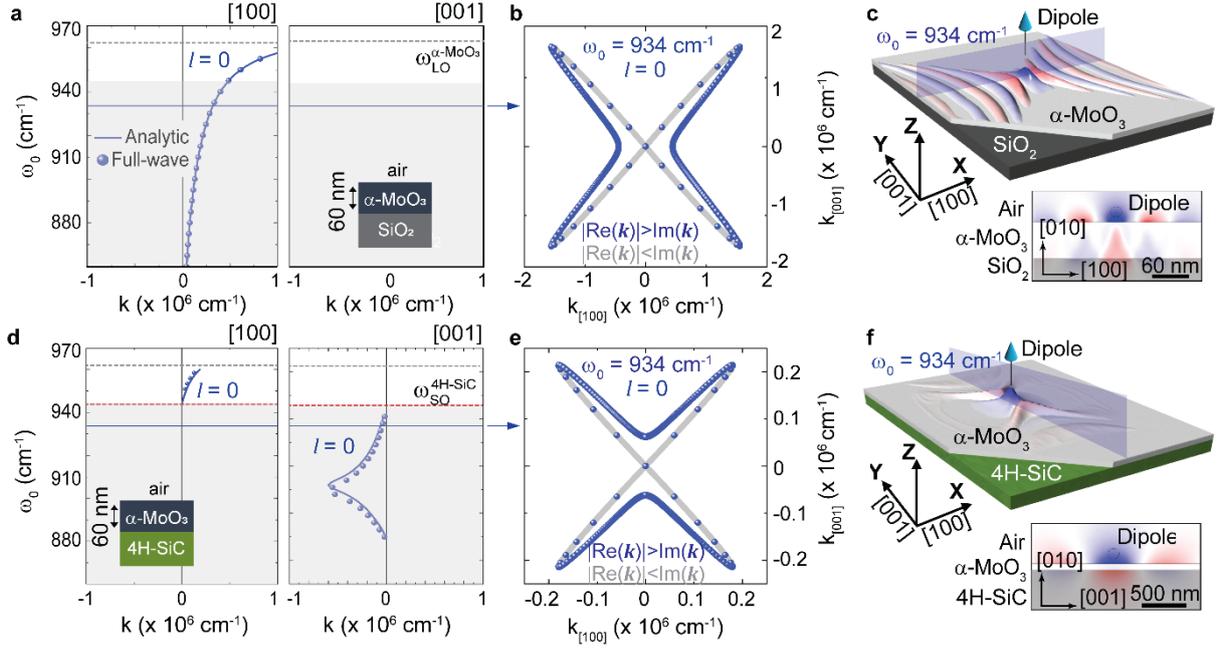

**Fig. 1 | Rotation of the directional propagation of in-plane HPhPs in an α-MoO₃ biaxial slab. a,** Dispersion of HPhPs in a 60-nm-thick α-MoO₃ slab on top of SiO₂ along both in plane-directions. The dashed grey line shows the LO phonon, while the continuous blue line marks $\omega_0 = 934$ cm⁻¹. **b,** IFC of HPhPs propagating in the α-MoO₃/SiO₂ heterostructure at an incident frequency $\omega_0 = 934$ cm⁻¹: the major axis of the hyperbola-like curve lays along $k_{[100]}$. **c,** Schematics of HPhPs propagating in the α-MoO₃/SiO₂ heterostructure, launched by a dipole source at $\omega_0 = 934$ cm⁻¹. Propagation occurs within hyperbolas centered along the [100] α-MoO₃ crystal direction. The inset shows a cross-section of the simulated electric-field distribution Re($E_z$), revealing that HPhPs in α-MoO₃/SiO₂ are volume- or slab-confined. **d,** Dispersion of HPhPs in a 60-nm-thick α-MoO₃ slab on top of 4H-SiC along both in plane-directions. The dashed grey line shows the LO phonon while the shadowed area indicates the area below the SO phonon of 4H-SiC (dashed red line). The continuous blue line marks $\omega_0 = 934$ cm⁻¹. **e,** IFCs of HPhPs propagating in the α-MoO₃/4H-SiC heterostructure at an incident frequency $\omega_0 = 934$ cm⁻¹, showing available wavevectors along $k_{[001]}$. **f,** Schematics of PhPs propagating in the α-MoO₃/4H-SiC heterostructure, launched by a dipole source at $\omega_0 = 934$ cm⁻¹. Propagation occurs within hyperbolas centered along the [001] α-MoO₃ crystal direction, in stark contrast with that in (C). The inset shows a cross-section of the simulated electric-field distribution Re($E_z$), revealing that HPhPs in α-MoO₃/4H-SiC are surface-confined.

To experimentally demonstrate this unique rotation of the directional propagation of HPhPs, we assembled both α-MoO₃/4H-SiC and α-MoO₃/SiO₂ heterostructures (see Methods) and compared the propagation of polaritons excited within them. Particularly, we performed real-space infrared nanoimaging (by employing scattering-type near-field optical microscopy, s-SNOM) of polaritons that are efficiently launched by a metallic antenna[26,31] fabricated on top of the heterostructures (see schematics in Fig. 2a, and Methods). We acquired near-field amplitude images at an incident laser frequency $\omega_O = 934$ cm⁻¹, since our theoretical calculations predict that, at this frequency, HPhPs will propagate along orthogonal in-plane directions in the two material heterostructures. For the α-MoO₃/SiO₂ heterostructure, our images show HPhPs propagating with concave wavefronts within hyperbolic sectors centered along the [100] direction, consistently with prior reports[16,17] (Fig. 2b, top panel). However, in the α-MoO₃/4H-SiC polaritonic heterostructure (Fig 2c), we observe HPhPs propagating with concave wavefronts within hyperbolic sectors centered along the [001] direction, which has thus far been a forbidden polaritonic direction in bare α-MoO₃ in this frequency range[16,17].

To further validate these experimental findings, we perform full-wave numerical simulations, where instead of a metallic antenna, a vertically oriented electric dipole placed on top of the heterostructures acts as the polaritonic source (Methods). The obtained near-field simulations show HPhPs propagating along the [100] in-plane direction in the α-MoO$_3$/SiO$_2$ heterostructure (Fig. 2b, bottom panel) and along its [001] orthogonal direction in the α-MoO$_3$/4H-SiC heterostructure (Fig. 2c, bottom panel). Taken together, our results unambiguously uncover a 90º rotation of the major axis of the polaritonic IFC, corroborating that the directional propagation of in-plane HPhPs in a biaxial slab can be effectively engineered along previously forbidden directions, as predicted by our analytical study. Note that we choose α-MoO$_3$ flakes of different thicknesses in both heterostructures (360 and 61 nm for α-MoO$_3$/SiO$_2$ and α-MoO3/4H-SiC, respectively) so that the wavelengths of HPhPs supported by them are similar, thus facilitating the comparison of the IFCs.

Further insights into the rotation of the IFC can be obtained by extracting the HPhPs dispersion in each of the heterostructures from near-field measurements at different incident frequencies $\omega_O$, and comparing them to the imaginary part of the Fresnel reflection coefficient, Im($r_p$), obtained from transfer-matrix calculations[32] (Figs. 2d, and 2e). As expected, the calculated polaritonic dispersion in α-MoO$_3$/SiO$_2$ (visible as bright maxima in the color plot) appears only along the [100] crystal direction[16,17,28] (Fig. 2d). In turn, the α-MoO$_3$/4H-SiC heterostructure shows polaritonic branches along both the [100] crystal direction (Fig. 2e), above $\omega_{SO}$, and [001] crystal direction, below $\omega_{SO}$, i.e. when the permittivity of 4H-SiC takes values below -1, in agreement with our analytical theory (Fig. 1). Besides, as illustrated in Fig. 1d, the polaritons exhibit negative phase velocity $v_p = \omega/k$, and, remarkably, a very flat dispersion curve[33,34] yielding ultraslow group velocities of about $10^{-5}$c (Supplementary Information).

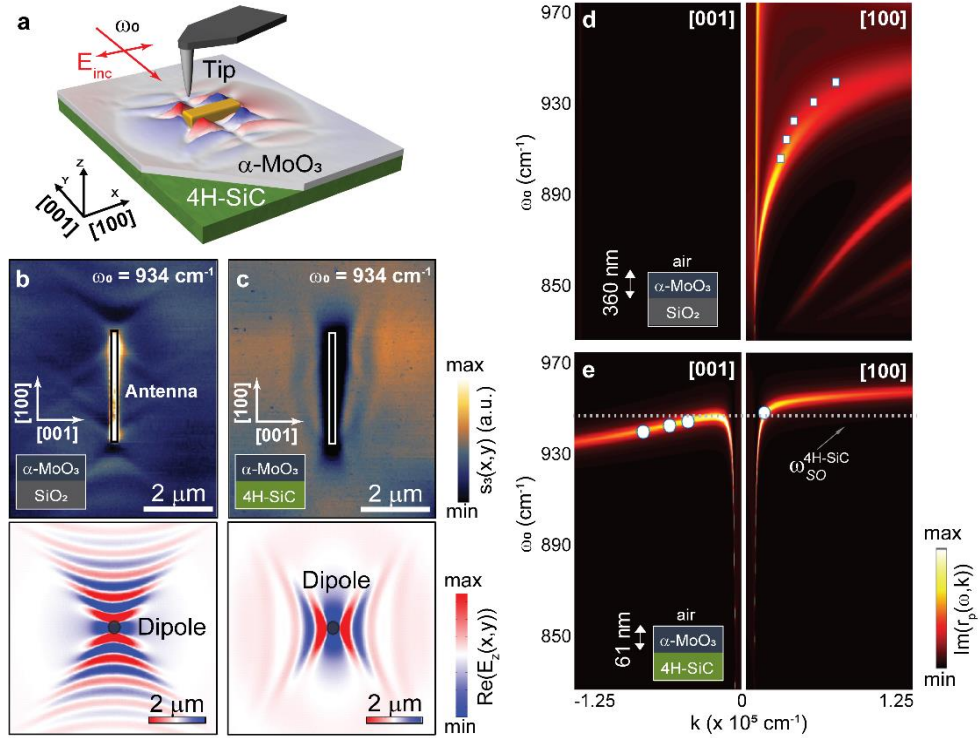

**Fig. 2 | Visualization of the rotation of the directional propagation of in-plane HPhPs in an α-MoO₃/4H-SiC heterostructure. a,** Illustration of antenna-launched HPhPs in an α-MoO$_3$/4H-SiC heterostructure. The spatial distribution of the near-field (shown in red and blue) is adapted from the simulation of the near-field distribution, Re($E_z(x,y)$). **b, top panel,** Experimental near-field amplitude images $s_3(x,y)$ of antenna-launched HPhPs on a 360-nm-thick α-MoO$_3$ flake on top of SiO$_2$ at an incident frequency $\omega_0 = 934$ cm$^{-1}$.

Propagation occurs within hyperbolic sectors centered along the [100] α-MoO$_3$ crystal direction. **b, bottom panel,** Simulated near-field distribution, Re($E_z(x,y)$), excited by a point dipole on a 360-nm-thick α-MoO$_3$ flake on top of SiO$_2$ at $\omega_0$ = 934 cm$^{-1}$. **c, top panel,** Experimental near-field amplitude images $s_3(x,y)$ of antenna-launched HPhPs on a 61-nm-thick α-MoO$_3$ flake on top of 4H-SiC at $\omega_0$ = 934 cm$^{-1}$. Propagation occurs within hyperbolic sectors centered along the [001] α-MoO$_3$ crystal direction. **c, bottom panel,** Simulated near-field distribution, Re($E_z(x,y)$), excited by a point dipole on a 61-nm-thick α-MoO$_3$ flake on top of 4H-SiC at $\omega_0$ = 934 cm$^{-1}$. **d,** Transfer-matrix calculations (false color plot) of HPhPs propagating on a 360-nm-thick α-MoO$_3$ flake on top of SiO$_2$ along the [100] (left panel) and [001] (right panel) crystal directions. White dots indicate experimental data from monochromatic s-SNOM imaging. **e,** Transfer-matrix calculations (false color plot) of HPhPs propagating on a 61-nm-thick α-MoO$_3$ flake on top of 4H-SiC along the [100] (left panel) and [001] (right panel) crystal directions. White dots indicate experimental data from monochromatic s-SNOM imaging. The dashed white line marks the SO phonon of 4H-SiC at $\omega_{SO}$ = 943 cm$^{-1}$.

To gain insights into the transition occurring between such mutually rotated propagating regimes, we image antenna-launched polaritons in an α-MoO$_3$/4H-SiC heterostructure at several intermediate frequencies between $\omega_O$ = 948 cm$^{-1}$ and $\omega_O$ = 937 cm$^{-1}$, i.e. in the frequency range in which the rotation takes place. At the highest frequency, $\omega_O$ = 948 cm$^{-1}$, we observe HPhPs propagating within a hyperbolic sector whose major axis lays along the [100] α-MoO$_3$ direction (Fig. 3a). However, when lowering the frequency to $\omega_O$ = 943 cm$^{-1}$, propagation of HPhPs takes place also within a hyperbolic sector whose major axis lays along the [001] direction (Fig. 3b). This behaviour becomes even more evident at the slightly lower frequency $\omega_O$ = 941 cm$^{-1}$ (Fig. 3c), where exotic propagation of polaritons within hyperbolic sectors whose major axes lay along both orthogonal directions is clearly seen. Finally, by further lowering the frequency to $\omega_O$ = 937 cm$^{-1}$, propagation of polaritons is only observed within a hyperbolic sector whose major axis lays along the [001] crystal direction (Fig. 3d), thus completing a 90º rotation from the initial directional propagation of HPhPs at $\omega_O$ = 948 cm$^{-1}$.

We perform additional theoretical analyses to understand these unusual real-space images of propagating polaritons. Namely, we numerically calculate the near-field distribution Re($E_z(x,y)$) generated at the aforementioned frequencies by a localized source placed above a 61-nm-thick α-MoO$_3$ flake on top of 4H-SiC, and subsequently, we Fourier-transform (FT) them (left panels of Figs. 3e-h). The resulting FTs clearly show the rotation of the major axes of the hyperbolic IFCs from the [100] to the [001] α-MoO$_3$ crystal directions when comparing the extreme cases at $\omega_O$ = 948 and 937 cm$^{-1}$. However, apart from the rotation effect, at the intermediate frequencies $\omega_O$ = 943 and 941 cm$^{-1}$, the FTs also show non-trivial features appearing at relatively low momenta (Figs. 3f,g). Although these features point out to a complicated multimode excitation in this frequency range, by performing a careful analysis of the generated field patterns based on the Green's dyadic and simplified dispersion relations (see Supplementary Information), we find that only two modes excited in the heterostructure present a major contribution to the fields launched by a localized source: the lowest-order mode ($l$ = 0, according to Eq. (1)) and the mode arising under the limit of a vanishing slab thickness $d \to 0$ (see Methods). In fact, by deriving the analytical IFCs of these two modes, characterized by a hyperbolic curve becoming linear and intersecting in one point at one specific frequency for the $l$ = 0 mode (blue curve in the right of Fig. 3f), and by the presence of available wavevectors along all in-plane directions with small modulus for the $d \to 0$ mode (red curve in the right of Fig. 3f), we can explain well the main non-trivial features in the FTs at intermediate frequencies (see Supplementary Information).

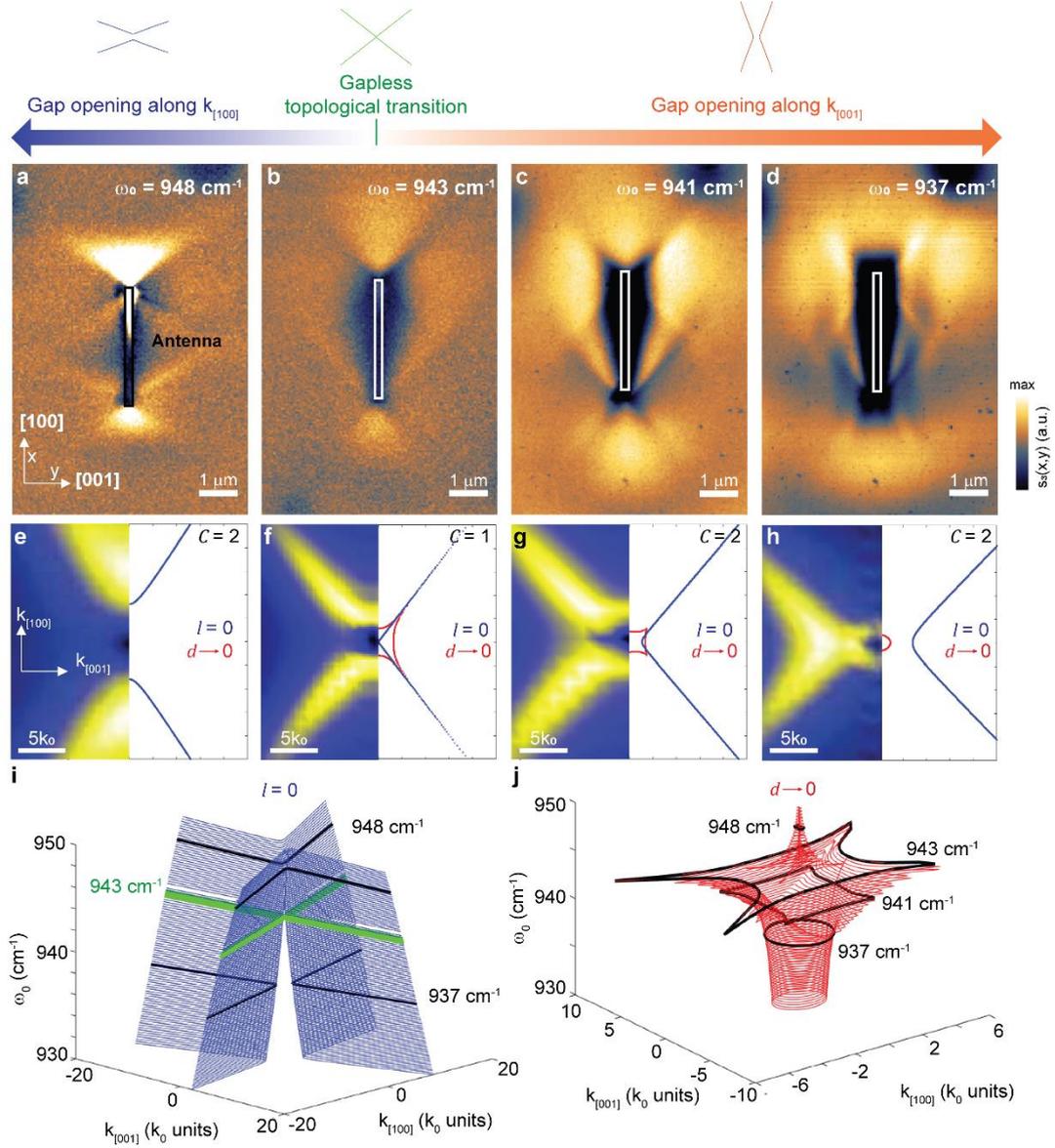

**Fig. 3 | Topological transition of the polaritonic IFC between orthogonal hyperbolic regimes in an α-MoO$_3$/4H-SiC heterostructure. a-c,** Experimental near-field amplitude images $s_3(x,y)$ of antenna-launched hyperbolic PhPs on a 61-nm-thick α-MoO$_3$ flake on top of 4H-SiC at different illuminating frequencies. **e-h, left,** Simulated IFCs, calculated by Fourier-transforming simulated near-field distributions, Re($E_z(x,y)$) created by a point dipole on a 61-nm-thick α-MoO$_3$ flake on top of 4H-SiC. **e-h, right** Analytical IFCs for the fundamental mode $l=0$ and the mode $d \to 0$. **i-j,** Three-dimensional visualization of the topological transition of the fundamental mode $l=0$ and of the mode $d \to 0$, in blue and red, respectively. Black lines show the IFCs at the frequencies at which the near-field images were acquired. The green line depicts the IFC at the topological transition.

Interestingly, it is worth having a look at the path-connectedness —i.e. the number of components in which a path can be drawn within any two points of a curve— of the analytical IFCs for the $l=0$ mode, as it is one of the main properties that are used to distinguish different topological spaces. Unlike size or shape, path-connectedness remains unchanged under homeomorphisms, and thus based on it we can distinguish when two curves are not topologically equivalent, i.e. not homeomorphic. As such, at $\omega_0$

= 948 cm$^{-1}$, 941 cm$^{-1}$ and 937 cm$^{-1}$ the number of path-connected components $C$ of the IFC of the $l = 0$ mode is 2. However, at $\omega_O = 943$ cm$^{-1}$, the gap in the IFC strikingly closes, and the number of path-connected components of the IFC evolves from $C = 2$ to $C = 1$. This IFC gap opening/closing of the IFC corresponds to cutting/attaching the polaritonic dispersion in momentum space and alters the topology of the system[35], as it represents a discontinuous deformation in which the topological invariant of the polaritonic mode $l = 0$ abruptly changes (see Supplementary Information). This unveils one of the landmark features of the three-dimensional polaritonic topological transition that emerges for the IFC of the lowest-order polaritonic mode in the frequency range between both mutually rotated regimes in the α-MoO$_3$/4H-SiC heterostructure. Such transition can thus be more clearly visualized in a three-dimensional plot of the in-plane dispersion of the IFC as a function of frequency $\omega$, i.e. in the three-dimensional space $(k_x, k_y, \omega)$ (Fig. 3i). The resulting surface is a hyperbolic cylinder composed of two disconnected sheets with their in-plane major axes along the $k_{[001]}$ direction at low frequencies and along the $k_{[100]}$ direction at high frequencies, respectively. These sheets become connected at the transition frequency $\omega_O = 943$ cm$^{-1}$, forming a cross-like curve (depicted in green) and clearly illustrating the topological transition of the IFC. In Fig. 3j, we show a three-dimensional plot of the mode $d \to 0$ as a function of frequency $\omega$, a surface composed of closed IFCs with low momentum. The cross-sections of this surface dramatically change with frequency (from a circle-like to a four-vertex star-like IFC) although preserving the same topology up to $\omega_O = 943$ cm$^{-1}$, where the surface virtually collapses into a point. Thus, we attribute the peculiarities observed in near-field images in Fig. 3a-d to the strong dependence of the propagation direction of polaritons with frequency, manifesting itself as a robust change of the IFCs of both modes, particularly near the topological transition.

In conclusion, we report engineering of the directional propagation of low-loss HPhPs in a natural medium along previously forbidden directions by the emergence of an optical topological transition when the crystal lays on a substrate with a specific negative permittivity. Contrary to other expected polaritonic topological transitions, as those at the limits between *reststrahlen* bands[22], and thereby close to TO or LO phonons where losses are high, the topological transition here reported when α-MoO$_3$ is placed on 4H-SiC takes place at $\omega_{SO} = 943$ cm$^{-1}$, where losses are low and thus optimal visualization of its unique features can be carried out in real-space. In turn, our results reveal two different regimes in which HPhPs propagate along mutually orthogonal directions, including the naturally forbidden [001] crystal direction for polaritonic propagation in bare α-MoO$_3$, and exotic intermediate states that unveil the topological origin of the transition, where HPhPs propagate within hyperbolas centered along both the [001] and [100] α-MoO$_3$ directions. Our work opens up stimulating horizons for extensive and effective control of the propagation of hyperbolic polaritons at the nanoscale, as the possibility of fabricating other polaritonic heterostructures thanks to the growing palette of polaritonic vdW materials that sum up to other traditional polaritonic media such as SiC, aluminium nitride (AlN), or quartz. Furthermore, the recent introduction of new degrees of freedom, such as the twist angle[23-26], can enable further external control of optical topological transitions naturally emerging in polaritonic heterostructures. In this context, our findings promise new opportunities for developing mid-infrared and terahertz[36] optical devices and for manipulating the flow of light at the nanoscale with unprecedented degree of control. Moreover, our study provides fundamentally relevant new insights into the emergence of optical topological transitions in low-loss vdW crystals, as well as a route through which to translate recent developments in topological engineering in low-dimensional electronic materials to photonics and polaritonics.